\documentclass[
aps,%
12pt,%
final,%
notitlepage,%
oneside,%
onecolumn,%
nobibnotes,%
nofootinbib,% 
superscriptaddress,%
noshowpacs,%
centertags]%
{revtex4}
\usepackage[english]{babel}
\usepackage{xspace}
\usepackage{morefloats}
\usepackage{epsfig}
\usepackage{pstricks}
\usepackage{graphicx}
\usepackage{subfigure}
\usepackage{amsfonts,amsmath,amssymb,bm,float,footnpag}
\newcommand{\kLu}{\ensuremath{\lvert\kappa_{\rm tug}\rvert/\Lambda}\xspace}
\newcommand{\kLc}{\ensuremath{\lvert\kappa_{\rm tcg}\rvert/\Lambda}\xspace}
\newcount\colveccount
\newcommand*\colvec[1]{
        \global\colveccount#1
        \begin{pmatrix}
        \colvecnext
}
\def\colvecnext#1{
        #1
        \global\advance\colveccount-1
        \ifnum\colveccount>0
                \\
                \expandafter\colvecnext
        \else
                \end{pmatrix}
        \fi
}
\providecommand{\PAKz}{\ensuremath{\overline{K}^0}\xspace} % anti-K0 meson
\newcommand{\PKz}{\ensuremath{\mathrm{K^0}}}
\newcommand{\PDz}{\ensuremath{{D^0}}}
\providecommand{\PADz}{\ensuremath{\overline{D}^0}\xspace} % anti-D0 meson
\newcommand{\PBz}{\ensuremath{{{B}^0}}}
\providecommand{\PABz}{\ensuremath{\overline{B}^0}\xspace} % anti-B0 meson
\newcommand{\cPqb}{\ensuremath{b}} % b for b quark
\newcommand{\PK}{\ensuremath{{K}}}
\newcommand{\PD}{\ensuremath{{D}}}
\newcommand{\PB}{\ensuremath{{B}}}

\begin{document}
%test
\selectlanguage{english}

\title{Eligibility of EFT Approach to Search for tqg FCNC Phenomenon}

\author{\firstname{E.E.} \surname{Boos}} 
\email{boos@theory.sinp.msu.ru}
\affiliation{Skobeltsyn Institute of Nuclear Physics,
Lomonosov Moscow State University}
\author{\firstname{V.E.} \surname{Bunichev}}
\email{bunichev@theory.sinp.msu.ru}
\affiliation{Skobeltsyn Institute of Nuclear Physics,
Lomonosov Moscow State University} 
\author{\firstname{L.V.} \surname{Dudko }}
\email{lev.dudko@cern.ch}
\affiliation{Skobeltsyn Institute of Nuclear Physics,
Lomonosov Moscow State University}
\author{\firstname{M.A.} \surname{Perfilov}} 
\email{Maksim.Perfilov@cern.ch}
\affiliation{Skobeltsyn Institute of Nuclear Physics,
Lomonosov Moscow State University}
\author{\firstname{G.A.} \surname{Vorotnikov}} 
\email{georgii.vorotnikov@cern.ch}
\affiliation{Skobeltsyn Institute of Nuclear Physics,
Lomonosov Moscow State University}

\begin{abstract}
Abstract -- The Effective Field Theory (EFT) approach is widely used in the search for possible deviations from the predictions of the Standard Model. Such an approximation of possible BSM physics is valid up to a certain levels of energy scale and accuracy. In this article, we investigate potential  limitation of the EFT approach related to unitarity to describe possible contributions of flavor changing neutral currents (FCNC) involving the top quark. The numerical and analytical calculations of the FCNC processes used in the EFT approach demonstrate the constant asymptotic behavior of the total cross section with increasing energy. It is shown that the EFT approach for studying the possible contribution of FCNC does not violate the restrictions following from perturbative unitarity, the asymptotic behavior of the cross section does not exceed the Froissart bound, and the approach itself can be used to set the corresponding experimental limits for FCNC couplings or Wilson coefficients at present and  future colliders.
\end{abstract} 
\maketitle

\section{Introduction. The EFT Lagrangian}
Flavour-changing neutral currents (FCNC) are absent at lowest order in the SM, and are significantly suppressed through the Glashow--Iliopoulos--Maiani 
mechanism~\cite{Glashow:1970gm} at higher orders.
Various rare decays of  $\PK$, $\PD$, and $\PB$ mesons, as well as 
the oscillations in $\PKz\PAKz$, $\PDz\PADz$, and $\PBz\PABz$
systems, strongly constrain FCNC interactions involving the first two 
generations and the \cPqb\ quark~\cite{Agashe:2014kda}.
However, FCNC involving the top quark are significantly less constrained.
In the SM, the FCNC couplings of the top quark are predicted to be very 
small and not detectable at current experimental sensitivity.
However, they can be significantly enhanced in various SM extensions. 

The FCNC interactions of the top quark with the quarks from the first two generations can be encoded in an effective field theory through dimension six gauge-invariant
 operators as indicated using slightly different notations in~\cite{Buchmuller:1985jz}~(see, eq.~3.61) and  in~\cite{Malkawi:1995dm,AguilarSaavedra:2008zc,Grzadkowski:2010es,Willenbrock:2014bja,AguilarSaavedra:2018nen}.
 The gauge-invariant effective Lagrangian for the tug FCNC interactions has
 the following form in the notations~\cite{AguilarSaavedra:2018nen}:
\begin{eqnarray}
\label{effective-lgrn}
{\cal L_{\rm EFT}} \,= \,
\frac{C^{13}_{uG}}{\Lambda^2}\,
\left( \bar u_L \bar d_L\right)\,\sigma^{\mu\nu}\,t^a\,t_R\, \Phi^c\, 
G^a_{\mu\nu}\,+\,\rm h.c.\nonumber\\
 \,+\, 
\frac{C^{31}_{uG}}{\Lambda^2}\,
\left( \bar t_L \bar b_L\right) \,\sigma^{\mu\nu}\,t^a \,u_R \,\Phi^c \,
G^a_{\mu\nu}\,+\,\rm h.c. ,
\end{eqnarray}
where $\Lambda$ is the scale of new physics (${\gtrsim} 1$ TeV), 
$\left( \bar u_L \bar d_L\right)$ and $\left( \bar t_L \bar b_L\right)$ are the left quark doublets,  
$C^{13}_{uG}$ and $C^{31}_{uG}$ are Wilson coefficients being complex in general, $t^a$ are 
the generators of the SU(3) color gauge group, $G^{\rm a}_{\mu\nu}$ is a
 gluon field strength tensor, and $\Phi^c$ is the conjugated Higgs field doublet.
The effective Lagrangian  for the tcg FCNC interactions  has the same form as
 (\ref{effective-lgrn}) with
obvious replacement of the $u$ and $d$ quarks by $c$ and $s$ quarks, and
 $C^{13}_{uG}$ and $C^{31}_{uG}$ by $C^{23}_{cG}$ and $C^{32}_{cG}$ respectively.

In the unitary gauge the conjugated Higgs field doublet has a well known 
form 
$$ \Phi^c = \colvec{2}{\frac{v+h}{\sqrt{2}}}{0},$$
where $h$ is the Higgs boson field and $v$ is the Higgs vacuum expectation value.
It is easy to see that in the unitary gauge the Lagrangian (\ref{effective-lgrn}) 
gets the following form:
\begin{eqnarray}
\label{lgrn_in_unigauge}
{\cal L_{\rm EFT}}= \frac{v+h}{\sqrt{2}}\,\frac{1}{\Lambda^2}\,
\big[\,
\bar q \,\sigma^{\mu\nu}\,t^a \,
\big(\,C^{i3}_{qG}\,\frac{1+\gamma^5}{2}\,+\,(C^{3i}_{qG})^*
 \,\frac{1-\gamma^5}{2}\,\big)\, t  \nonumber \\
\,+\, \bar t\, \sigma^{\mu\nu}\,t^a
\big(\,(C^{i3}_{qG})^*\,\frac{1-\gamma^5}{2}\,+\,C^{3i}_{qG} 
\, \frac{1+\gamma^5}{2}\,\big)\, q \big]\, 
G^a_{\mu\nu}, 
\end{eqnarray}
where q refers to either $u$ ($i = 1$) or $c$ ($i = 2$) quarks.

In case of $C^{i3}_{uG}=(C^{3i}_{uG})^*$ 
the Lagrangian (\ref{lgrn_in_unigauge}) becomes:
\begin{eqnarray}
{\cal L_{\rm EFT}}= \frac{v+h}{\sqrt{2}}\,\,\frac{1}{\Lambda^2}\,
\big(\,C^{i3}_{qG}\,\bar q \,\sigma^{\mu\nu}\,t^a\, t 
 \,+ \,
(C^{i3}_{qG})^*\,\bar t\, \sigma^{\mu\nu}\,t^a\, q\,\big)\, G^a_{\mu\nu}.
\label{Lagrangian}
\end{eqnarray}
 One should note that 
the Lagrangian (\ref{Lagrangian})
contains not only the FCNC vertex of the top quark interaction with the u quark 
and gluon (tqg) but also the four-point FCNC vertex involving two 
gluons (tqgg). The latter is needed to preserve gauge invariance. The
Lagrangian (\ref{Lagrangian}) also includes the FCNC top quark interactions involving 
the Higgs boson
(tqgh) and (tqggh)\footnote[1]{The Lagrangian describing the tree point 
(tqg) and the four-point (tqgg) interaction vertices (tqg) as well as 
corresponding Feynman rules were presented in~\cite{Malkawi:1995dm}. The 
Lagrangian describing not only the tree point vertex (tqg), but also the 
interaction vertex with the Higgs boson (tqgh) as follows from the 
dimension 6 operators was worked out in~\cite{AguilarSaavedra:2008zc}.}.
but also the four-point FCNC vertex involving two gluons (tqgg). The latter is needed to preserve gauge invariance.
 The Lagrangian (\ref{Lagrangian}) also includes the FCNC top quark interactions involving the Higgs boson (tqgh) and (tqggh). But corresponding processes are 
obviously out of reach at the LHC and will be not considering. 
The experimental limits~\cite{Aad:2015gea,Khachatryan:2016sib,Aaltonen:2008qr,Abazov:2010qk} are given in terms of \kLu and \kLc couplings~\cite{Beneke:2000hk} which can be rewritten in the form of $C^{i3}_{qG}$ coefficients as follows:
\begin{eqnarray}
\label{couplings}
\lvert\kappa_{\rm tqg}\rvert/\Lambda = \frac{1}{g_s}\frac{v}{\sqrt{2}}\frac{C^{i3}_{qG}}{\Lambda^2},
\end{eqnarray}
where $g_s$  is the coupling constant of the strong interaction. 

\section{Numerical estimation of the FCNC tqg contributions }
\begin{figure}[t!]
\begin{center}
% \setcaptionmargin{-25mm}
\onelinecaptionsfalse
\includegraphics[width=.9\linewidth]{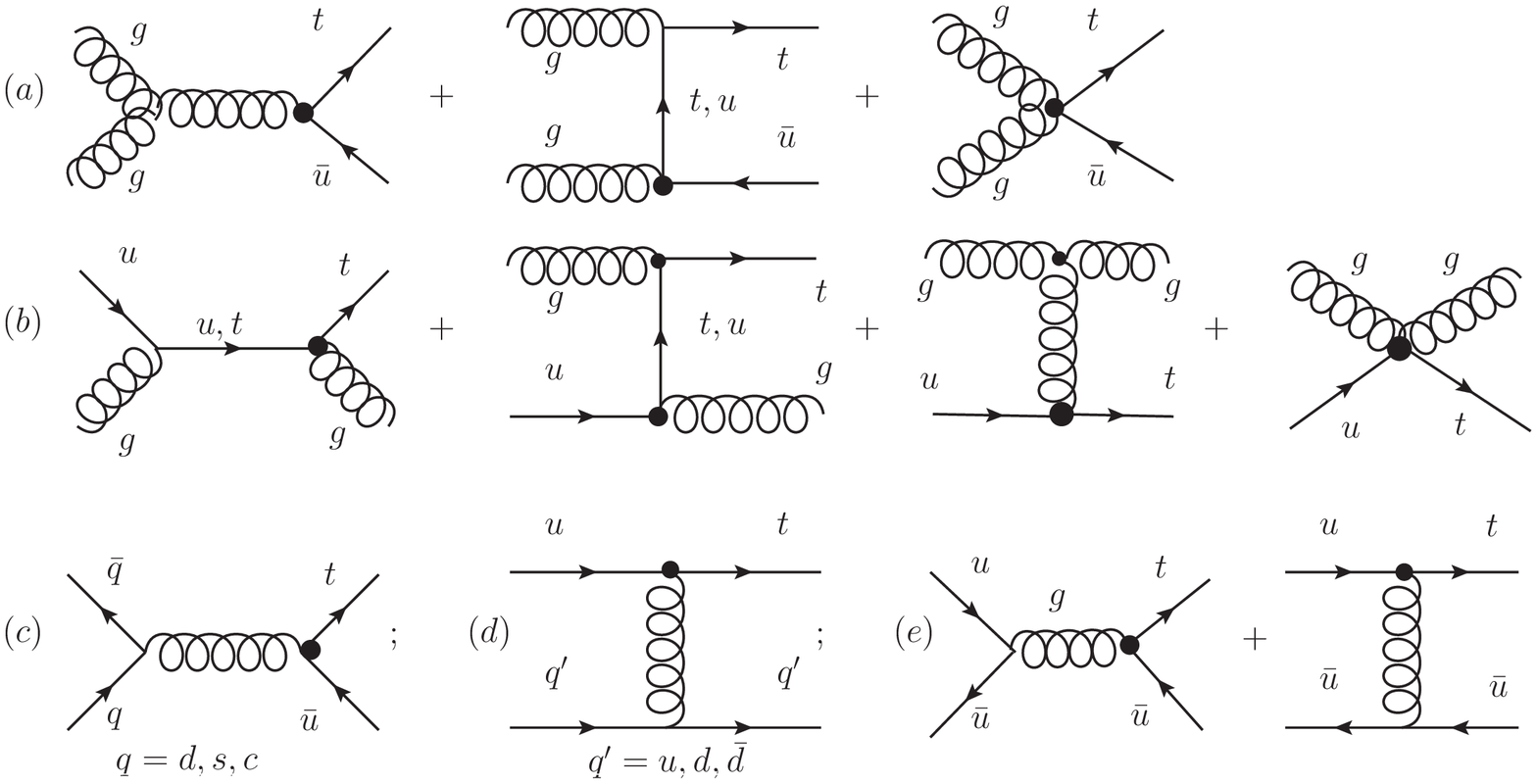} % Так вставляется рисунок
\captionstyle{normal}
\caption{Representative Feynman diagrams for the top quark production via tug FCNC interactions are described in eq.~(\ref{Lagrangian}).}
\label{diag}
\end{center}
\end{figure}

The representative set of Feynman diagrams for the top quark production in FCNC interactions described by Lagrangian~(\ref{Lagrangian}) are shown in Fig.~\ref{diag}. In this paper we consider $pp\to tj$ process, where j means a jet originated from either a light flavor quark or a gluon. Based on the Lagrangian~(\ref{Lagrangian}) necessary Feynman rules were implemented to the CompHEP~\cite{Pukhov:1999gg,Boos:2004kh} and all numerical calculations are performed by means of this package\footnote[2]{Note that the four-point vertex of the gluon--gluon--top~ quark--$u$~quark interaction from (\ref{Lagrangian}) was implemented in CompHEP using an auxiliary color octet field with spin two $t_{\mu\nu}^a$, denoted as $G.t$ in CompHEP, with the propagator defined by the Lagrangian $-\frac{1}{2}\,t_{\mu\nu}^a \, t^{\mu\nu}_a$. It should be noted that the four-point vertex is necessary for gauge invariance in calculating the contributions with the initial states $gg$ and $gu$.}. Calculations are performed in Feynman gauge.

The total cross section at 14 TeV collision energy is about 120 pb for some particular value of the 
parameter \kLu=0.03 TeV$^{-1}$ and requirements of transverse momenta and pseudorapidity of the final light flavor quark or gluon to be $P_T^{q,g}>20$ GeV, $|\eta^{q,g}|<4$. The cross section depends quadratically on the \kLu and can be recalculated for all other values of the coupling. 
The partial contribution of the processes with different initial states are 
$gg$ (2\%) (Fig.~\ref{diag}a), % 12
$qg$ (89\%) (Fig.~\ref{diag}b), % 5,11
$q\bar q$ (0.03\%) (Fig.~\ref{diag}c), % 7,10
$uq^\prime$ (7.8\%) (Fig.~\ref{diag}d) and % 1,2,4,6,9
$u\bar u$ (1.2\%) (Fig.~\ref{diag}e) % 3,8
for the LHC collider energy $\sqrt{s}=14$ TeV. 
The destructive contributions of the last diagrams with four-point vertex in~Figs.~\ref{diag}a,~\ref{diag}b decrease the cross section of $gg$ part by 60\% and $qg$ part by 4\%, while practically not affecting kinematic properties. 

\begin{figure}
\begin{center}
% \setcaptionmargin{-25mm}
\onelinecaptionstrue
\includegraphics[ width=.7\linewidth]{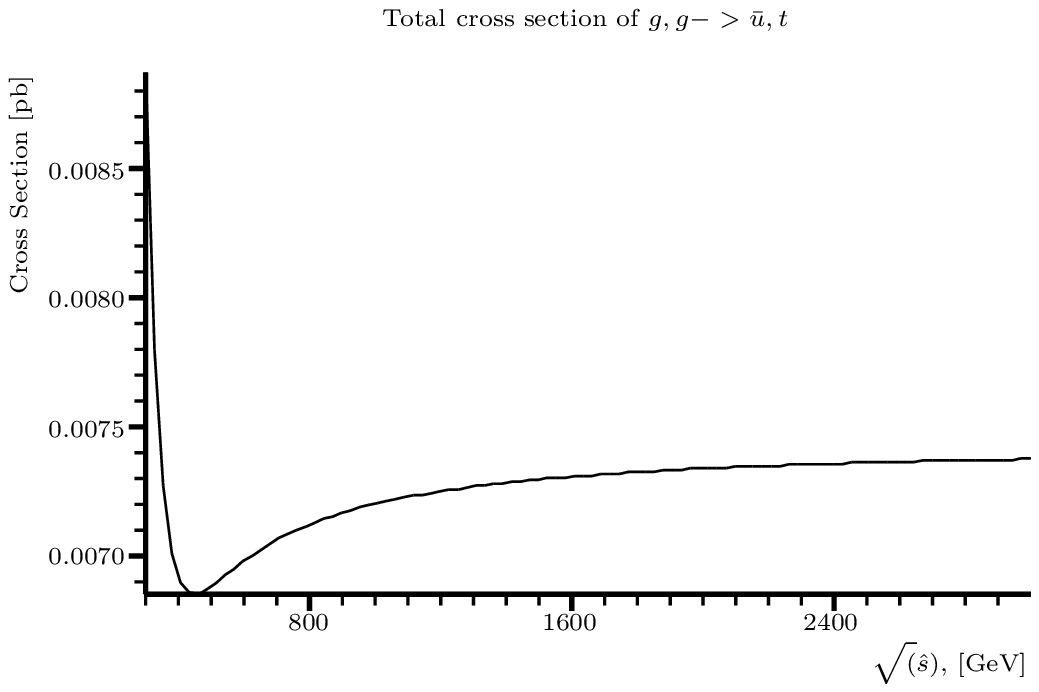} % Так вставляется рисунок
\captionstyle{normal}
\caption{Dependence of the total cross section of parton level FCNC process with $gg$ initial states (Fig.~\ref{diag}a) on the $\sqrt{\hat s}$. 
The region $300<\sqrt{\hat s}<3000$ GeV is shown.}
\label{gg-shat}
\end{center}
\end{figure}

\begin{figure}
\begin{center}
% \setcaptionmargin{-25mm}
\onelinecaptionstrue
\includegraphics[ width=.7\linewidth]{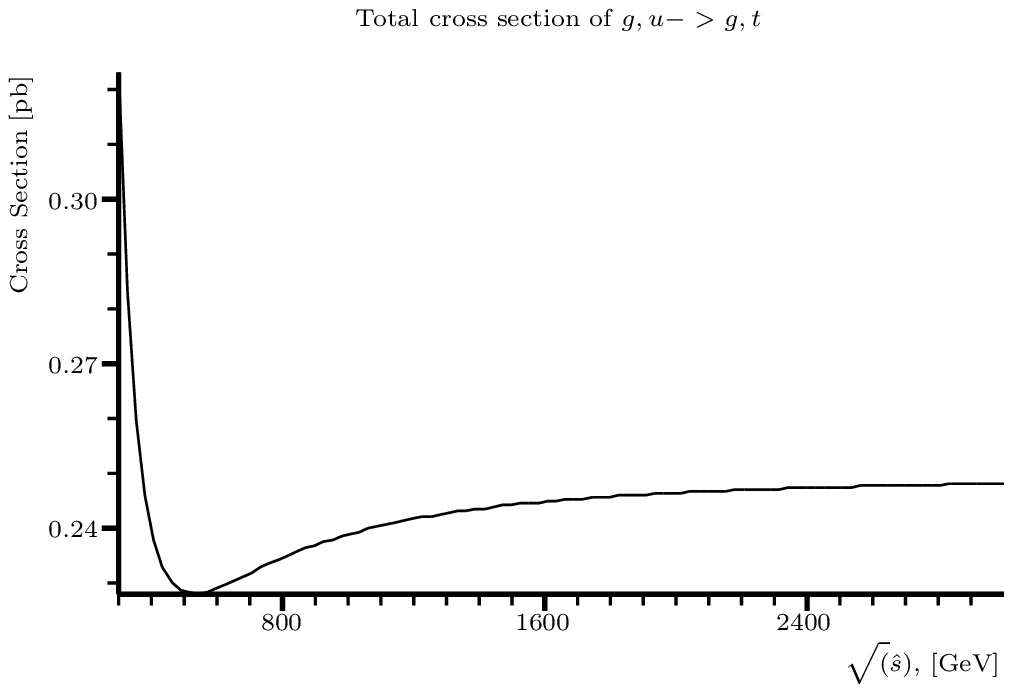} % Так вставляется рисунок
\captionstyle{normal}
\caption{Dependence of the total cross section of parton level FCNC process with $gu$ initial states (Fig.~\ref{diag}b) on the $\sqrt{\hat s}$. 
The region $300<\sqrt{\hat s}<3000$ GeV is shown.}
\label{gu-shat}
\end{center}
\end{figure}

\begin{figure}
\begin{center}
% \setcaptionmargin{-25mm}
\onelinecaptionstrue
\includegraphics[ width=.7\linewidth]{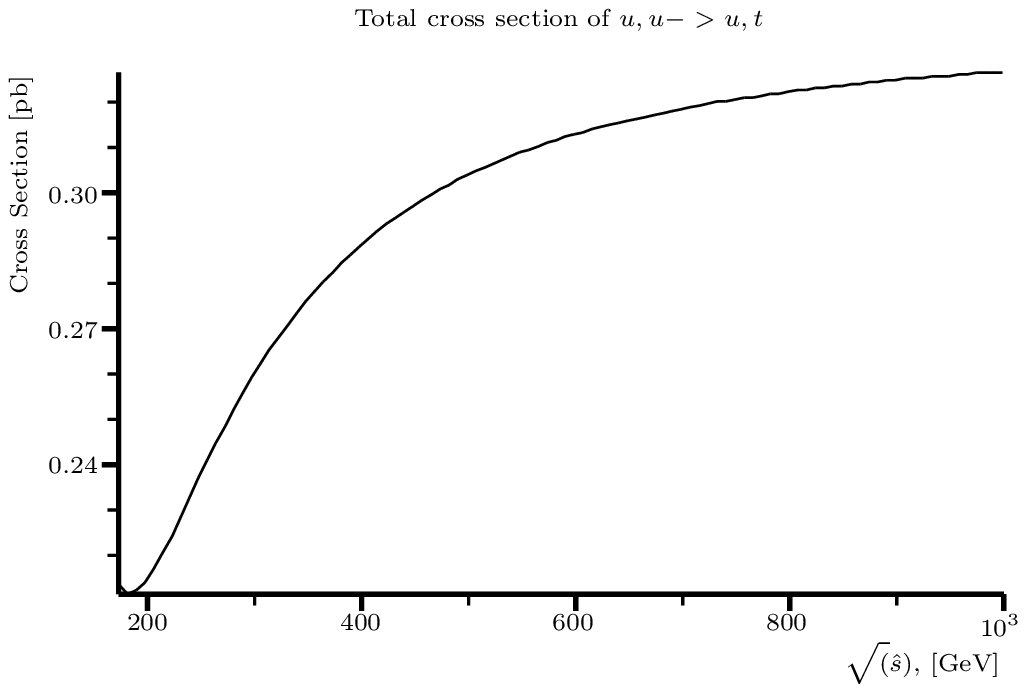} % Так вставляется рисунок
\captionstyle{normal}
\caption{Dependence of the total cross section of parton level FCNC process with $uu$ initial states (Fig.~\ref{diag}d) on the $\sqrt{\hat s}$.
The region $174<\sqrt{\hat s}<1000$ GeV is shown.}
\label{uu-shat}
\end{center}
\end{figure}

In order to analyze properties related to unitarity we need to check the behavior of FCNC contribution, introduced in EFT approach, with increasing the $\hat s$. The most transparent way is to exclude convolution with Proton Density Functions (PDF) and check the dependence of the total cross section of the parton level processes on the $\sqrt{\hat s}$. The Fig.~\ref{gg-shat} demonstrates dependence of the total cross section of the parton level process with $gg$ initial states (Fig.~\ref{diag}a) on the $\sqrt{\hat s}$, without integration over PDF. The same dependencies are shown in Fig.~\ref{gu-shat} for the $gu$ initial state (Fig.~\ref{diag}b) and in Fig.~\ref{uu-shat} for the $uu$ initial state (Fig.~\ref{diag}d). The required cutoff was applied as $|\cos (p_i,p_f)|<0.95$, where $p_i$ and $p_f$ are momenta of the initial and final partons. The wide region of possible $\sqrt{\hat s}$ was tested, started from the threshold about 173 GeV and up to the 100 TeV. For the better clarity only the region with visible changes are shown in Figs.~\ref{gg-shat}--\ref{uu-shat} and the cross section becomes constant with increasing the collision energy and without convolution with PDF. The overall behavior demonstrates the absence of the increasing of the cross section with the energy up to the 100 TeV of $\sqrt{\hat s}$. This observation is confirmed by the direct symbolic computation of asymptotic behavior of the cross section at large $\sqrt{\hat s}$ which is demonstrated in the next section.

\section{Unitary limit }

Effective operators lead to growing contributions with energies that violate unitarity. In order for our calculations to be self-consistent, we have to check that we do not consider kinematic regions where perturbative unitarity is violated. To estimate the allowed region of parameters, we apply optical theorem, which follows from the unitarity of the S-matrix. The optical theorem says that the imaginary part of the forward scattering amplitude is proportional to the total cross section of the process.
\begin{align}\label{sigma1}
\sigma = \frac{1}{s}{\rm Im}\left(A(\theta=0)\right)=\frac{16\pi}{s}\sum\limits^{\infty}_{l=0}(2l+1)|a_l|^2,
\end{align}
where $a_l$ is the partial-wave amplitude. Hence, Im$a_l=|a_l|^2$, which means that: 
\begin{align}\label{unitarity}
|a_0| < \frac{1}{2}.
\end{align}
The complete expression for the amplitude of the process $uu\to ut$ has the form:

\begin{align}\label{amplitude1}
A = 2g_s^2\frac{\lvert\kappa_{\rm tqg}\rvert}{\Lambda}\cdot\sum\limits_{spin}&~\big[~\bar{u}_u(p_3)(-i  t^a\gamma^{\mu})u_u(p_1)\left(\frac{-i g_{\mu\nu}}{(p_4-p_2)^2}\right)\bar{u}_t(p_4)\left(t^a\sigma^{\nu k}(p_4-p_2)_k\right) u_u(p_2)\\
\nonumber &
+\bar{u}_t(p_4)\left(t^a\sigma^{\mu k}(p_4-p_1)_k\right) u_u(p_1)\left(\frac{-i g_{\mu\nu}}{(p_4-p_1)^2}\right)\bar{u}_u(p_3)(-i t^a\gamma^{\nu})u_u(p_2)~\big].
\end{align}

After the convolution of Lorentz indices and summation over all spin states, as well as after the expression of scalar products via Mandelstam variables, the amplitude takes the form:
\begin{align}\label{amplitude2}
A = 8\cdot g_s^2\cdot \frac{\lvert\kappa_{\rm tqg}\rvert}{\Lambda}\cdot\sqrt{s\cdot t\cdot u}\left(\frac{1}{t} + \frac{1}{u}\right).
\end{align}
Here we assume that $\sqrt{s}\gg M_{\rm top}$.
Using the substitution $(u = -s-t)$ and integrating over the variable $t$, we obtain the partial-wave amplitude $a_0$:
\begin{align}\label{a0}
|a_0| = \frac{1}{16\pi s}\left| \int\limits^{0}_{-s}dt\cdot A \right | = 2\pi\alpha_s\cdot \frac{\lvert\kappa_{\rm tqg}\rvert}{\Lambda}\cdot\sqrt{s}.
\end{align}
The unitarity condition, following from the optical theorem ($|a_0|<\frac{1}{2}$)
and recent experimental limits ($\kLu\leqslant 0.004\ \rm TeV^{-1}$~\cite{Khachatryan:2016sib}, $\alpha_s\approx 0.1$~\cite{Agashe:2014kda}) on FCNC coupling parameter leads to the following estimation:
\begin{align}\label{estimation}
\sqrt{s} < \frac{1}{4\pi\alpha_s}\cdot\frac{\Lambda}{\lvert\kappa_{\rm tqg}\rvert} \approx 200~\rm TeV.
\end{align}
Thus our estimation shows that the current limit value of FCNC coupling does not violate the unitary behavior of the amplitudes of the studied processes at energies available at the present and future colliders.

Performing the calculation of the matrix element square, averaging over the initial spin and color 
states, and integrating over the $\cos\theta$ in the range from 
($-1+\epsilon$) to ($1-\epsilon$), we obtain the following expression 
for the cross section of the process $uu\to ut$
in the limit of $\sqrt{s}\gg M_{\rm top}$ as a function of the energy 
$\sqrt{s}$ and the cutoff parameter $\epsilon$:
\begin{align}\label{estimation2}
\sigma(\epsilon) = \frac{64}{9}\pi\alpha_s^2\cdot\frac{\kappa^2_{\rm 
tqg}}{\Lambda^2}\cdot\left(\ln\left|\frac{2}{\epsilon}-1\right|-\frac{11}{12}(1-\epsilon)\right).
\end{align}
Formula (\ref{estimation2}) explicitly demonstrates
the constant asymptotic behavior at large $\sqrt{s}$ in case of the 
scattering angular cut is applied.

If one integrates the matrix element square over the transverse momentum 
of the t-quark in the range from the cut parameter $\delta p_T$ to the 
kinematic limit $\sqrt{s}/2$ one gets for the cross section at 
$\sqrt{s}\gg M_{top}$:
\begin{align}\label{estimation3}
\sigma(\delta p_T,s) = \frac{64}{9}\pi\alpha_s^2\cdot\frac{\kappa^2_{\rm 
tqg}}{\Lambda^2}\cdot\left(2\cdot\ln\left|\frac{\sqrt{s}}{2\cdot\delta 
p_T}\left(1+\beta\right)\right|-\frac{11}{12}\beta\right),
\end{align}
where: $\beta=\sqrt{1-\frac{4\cdot\delta p_T^2}{s}}$.

The formula (\ref{estimation3}) shows the logarithmic dependence of the
cross section at high energies
$\ln\left|\frac{\sqrt{s}}{\delta p_T}\right|$
in case of a cut on $p_T$.

This behavior of the scattering cross section is confirmed by numerical simulations performed in the previous section and is in a good agreement with the Froissart bound~\cite{Froissart:1961ux} (the cross section does not increase faster than the square of the logarithm of energy), which confirms the correctness of the investigated approach.

\section{Conclusion }
In the paper we investigate limitation of EFT approach to describe possible FCNC processes. Numerical and analytic calculations demonstrate constant asymptotic behavior of the total cross section with the increasing of the energy. It is shown in the paper, that the EFT approach for the introducing of the possible FCNC contribution does not violate the pertubative unitary limit and  Froissart bound, and can be used to setup the corresponding experimental limits on the FCNC couplings or Wilson coefficients at the present~\cite{Aad:2015gea,Khachatryan:2016sib} and future~\cite{CMS:2018kwi,CYRM950,Dudko:2643278,Abada:2019lih} colliders.

\section*{Acknowledgements}
\label{sec:acknownlegements}

The work was supported by grant no.~16-12-10280 of the Russian Science Foundation.

\section*{References}
\nocite{*}
\bibliography{eft_fcnc}
% \newpage
% \begin{itemize}
% \item Fig~\ref{diag}. Representative Feynman diagrams for the top quark production via tug FCNC interactions are described in eq.~(\ref{Lagrangian}).
% \item Fig~\ref{gg-shat}. Dependence of the total cross section of parton level FCNC process with $gg$ initial states (Fig.~\ref{diag}a) on the $\sqrt{\hat s}$. The region $300<\sqrt{\hat s}<3000$ GeV is shown.
% \item Fig~\ref{gu-shat}. Dependence of the total cross section of parton level FCNC process with $gu$ initial states (Fig.~\ref{diag}b) on the $\sqrt{\hat s}$. The region $300<\sqrt{\hat s}<3000$ GeV is shown.
% \item Fig~\ref{uu-shat}. Dependence of the total cross section of parton level FCNC process with $uu$ initial states (Fig.~\ref{diag}d) on the $\sqrt{\hat s}$. The region $174<\sqrt{\hat s}<1000$ GeV is shown.
% \end{itemize}
\newpage

\newpage
\end{document}